\documentclass[aps,prb,twocolumn,superscriptaddress,nofootinbib]{revtex4-2}

\usepackage{amsmath}
\usepackage{amssymb}
\usepackage{graphicx}
\usepackage{hyperref}
\usepackage{xcolor}

\begin{document}

\title{Photon echo and fractional excitation lensing of $S=1/2$ XY spin chain}
\date{\today}

\author{Zi-Long Li}
\affiliation{Institute of Physics, Chinese Academy of Sciences, Beijing 100190, China}
\affiliation{University of Chinese Academy of Sciences, Beijing 100049, China}

\author{Yuan Wan}
\email{yuan.wan@iphy.ac.cn}
\affiliation{Institute of Physics, Chinese Academy of Sciences, Beijing 100190, China}
\affiliation{University of Chinese Academy of Sciences, Beijing 100049, China}
\affiliation{Songshan Lake Materials Laboratory, Dongguan, Guangdong 523808, China}

\begin{abstract}
We study numerically the two-dimensional coherent spectrum (2DCS) of the Tomonaga-Luttinger liquid hosted by the $S=1/2$ XY spin chain. The 2DCS characterizes the system's third order nonlinear magnetic response triggered by three pulses, separated successively by the delay time and the waiting time. It exhibits a photon echo signal resulting from a lensing process of the fractional excitations: A pair of photon-excited fractional excitations, initially moving apart, reverse their direction of motion, and annihilate each other. In the XY chain, the nonlinearity in the dispersion relation of the Jordan-Wigner fermions leads to the dispersion of the fractional excitation wave packets and thereby suppresses lensing. The magnitude of the echo signal decreases exponentially with increasing delay time. The decay rate scales with the temperature $T$ as $T^n$ at low temperature, where $n$ is the leading order of the Jordan-Wigner fermion dispersion, and as $T$ at high temperature. By contrast, as the waiting time increases, the magnitude of the echo signal saturates, reflecting the integrability of the system. Our results illustrate the effectiveness of the 2DCS in detecting subtle dynamical properties of optical excitations in spin chains. 
\end{abstract}

\maketitle

%%%%%%%%%%%%%%%%%%%%
% Introduction
%%%%%%%%%%%%%%%%%%%%
\section{Introduction \label{sec:intro}}

The latest emergence of terahertz two-dimensional coherent spectroscopy (THz 2DCS) offers a new lens on the rich dynamical phenomena in condensed matter~\cite{Woerner2013}. Being a time domain, nonlinear optical spectroscopy, the 2DCS triggers quantum interference processes of the optical excitations with successive phase coherent pulses, thereby accessing information that is usually unavailable to conventional, linear optical spectroscopies~\cite{Mukamel1995,Hamm2011}. Operating in the meV energy window, THz 2DCS is well positioned to study collective excitations in quantum materials. Experimentally, it has revealed a host of interesting phenomena in quantum wells~\cite{Woerner2013}, antiferromagnets~\cite{Lu2017}, electronic glasses~\cite{mahmood2021observation}, and superconductors~\cite{luo2022quantum,zhang2023revealing}. Meanwhile, theorists have predicted its potential utility for a wide range of systems from quantum spin liquids to topological insulators~\cite{Wan2019,Choi2020,Parameswaran2020,li2021photon,nandkishore2021spectroscopic,fava2021hydrodynamic,hart2023extracting,gao2023two,negahdari2023nonlinear,sim2023nonlinear,sim2023shedding,qiang2023probing}.

\begin{figure}
\includegraphics[width = \columnwidth]{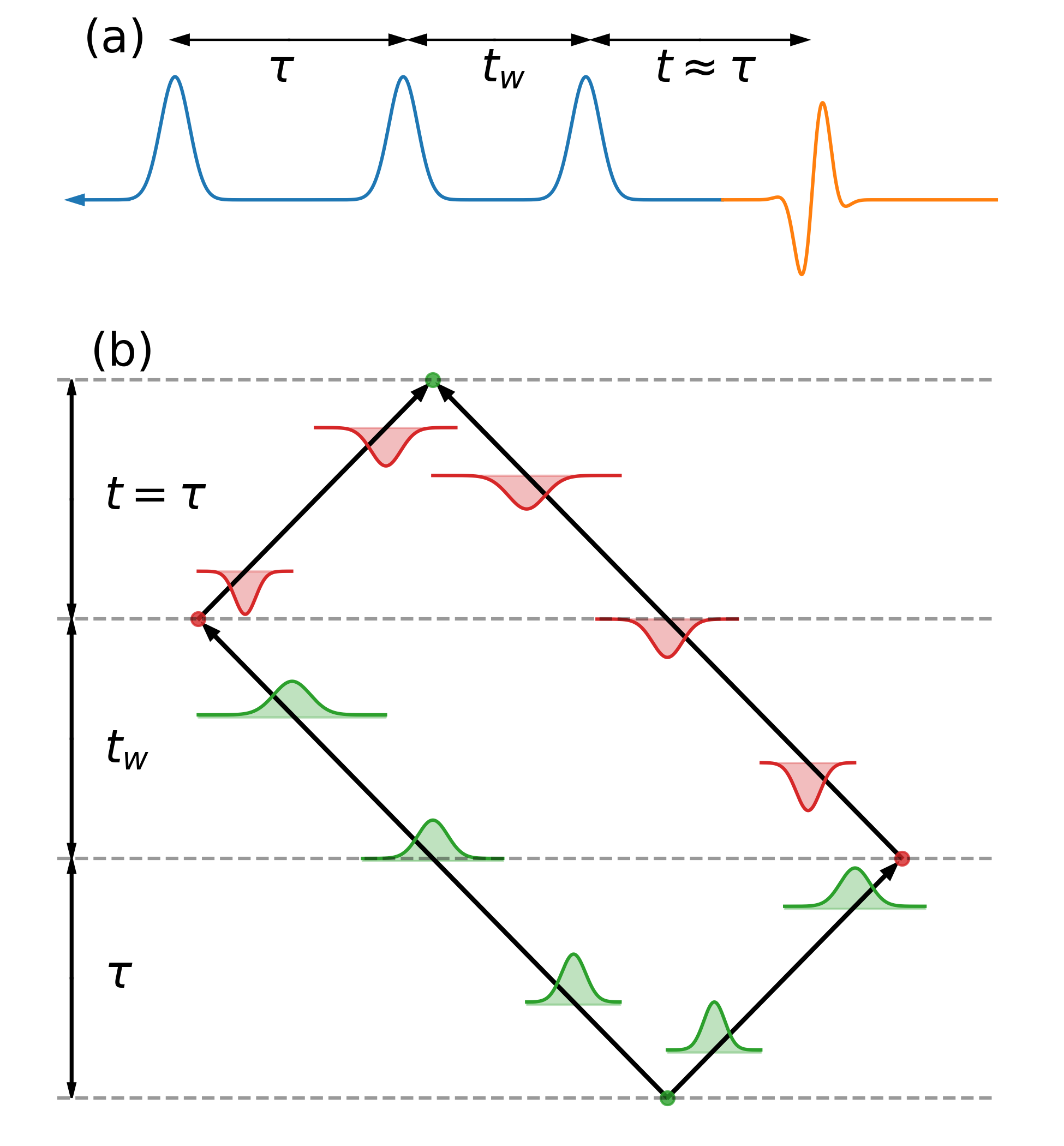}
\caption{(a) Photon echo is a third order nonlinear optical phenomenon triggered by three successive pulses, separated by the pulse delay time $\tau$ and the waiting time $t_w$. The echo (gold solid line) is the sudden rise of the response at a later time $t\approx \tau$. (b) The lensing of fractional excitations in the Tomonaga-Luttinger liquid (TLL) hosted by spin chains. In the case of the ferromagnetic XY chain ($J>0$ in Eq.~\eqref{eq:h_xy}, the first pulse creates a pair of spinons. In the second stage, the second and third pulses convert them to anti-spinons and reverse their directions of motion. Finally, the two anti-spinons annihilate and emit echo. In the case of the antiferromagnetic chain ($J<0$), the fractional excitations involved in the lensing process are a spinon and a Laughlin quasiparticle. Due to the dispersion of the wave packets, the second and the third pulses are no longer effective in refocusing the excitations' world lines. As a result, the lensing is suppressed.}
\label{fig:sche}
\end{figure}

A prominent feature of the 2DCS is its ability to measure the \emph{photon echo}~\cite{Kurnit1964}. Photon echo is a third-order nonlinear optical response initiated by three optical pulses,  separated successively on the time axis by the delay time $\tau$ and the waiting time $t_w$ (Fig.~\ref{fig:sche}(a)). After the arrival of the last pulse, the echo appears as the sudden rise of the nonlinear response at a later time $t\approx \tau$. Closely analogous to the spin echo~\cite{Hahn1950} in nuclear magnetic resonance, the photon echo is a sensitive probe for dissipation. Taking few-body systems as example, the fading of the echo with an increasing delay time $\tau$ is a direct manifestation of decoherence, whereas the echo's falling off with increasing waiting time $t_w$ reflects depopulation. 

This unique property of photon echo is traced back to the quantum interference process that produces the echo. In the case of few-body systems, the photon echo arises from an evolution trajectory of the density matrix (known as the Liouville pathway) that executes an effective time reversal operation. Such an operation erases all effects from the unitary evolution, and, consequently, exposes the dissipation~\cite{Mukamel1995,Hamm2011}. 

The profound connection between the photon echo and the quantum interference is amply demonstrated, as well as enriched, by examining many-body systems. In a previous analysis, we consider the Tomonaga-Luttinger liquid (TLL)~\cite{li2021photon} hosted by quantum spin chains. We find that its nonlinear magnetic response features a photon echo similar to few-body systems. However, the physical mechanism responsible for the echo is quite different. In this system, the photon echo arises from \emph{lensing} (Fig.~\ref{fig:sche}(b)), a spacetime interference process of the fractional excitations in TLL such as spinons and Laughlin quasiparticles~\cite{Pham2000}. The first optical pulse creates a pair of fractional exctiations that are moving apart. Under the action of the second and the third pulses, they reverse their directions of motion, and heading toward each other. In the last stage, the annihilation of the two excitations produces the echo.

Akin to the interference of waves, the lensing requires the coherent propagation of the wave packets of fractional excitations. Basing on the bosonization technique, the previous analysis approximates the TLL as a system of non-interacting bosons with linear dispersion relation. In this highly idealized situation, the photon echo is found to be independent of both the delay time $\tau$ and the waiting time $t_w$, i.e. the echo does not fade away. It is expected that dissipation and/or dispersion of the fractional excitations, which must occur in any realistic, microscopic spin chains, suppress lensing, and, consequently, the photon echo. Yet, these effects have not been subjected to quantitative analysis by far.

In this work, we make a first pass at this problem by studying the photon echo from $S=1/2$ spin XY chain~\cite{Lieb1961}, which hosts a TLL with the Luttinger parameter $K=1$. Being equivalent to an ensemble of non-interacting Jordan-Wigner fermions~\cite{Jordan1928}, this model possesses no inherent dissipation. However, owing to the nonlinearity in the dispersion relation of the Jordan-Wigner fermions, wave packets of fractional excitations disperse as they propagate through the TLL. Meanwhile, the integrability of this model gives access to its long time dynamics. Therefore, the $S=1/2$ spin XY chain is an ideal platform for investigating the impact of dispersion on lensing and photon echo. 

Through numerical calculation, we verify that the dispersion of the fractional excitations does lead to the decay of the photon echo. On one hand, the magnitude of the echo signal decreases exponentially with increasing delay time $\tau$. When the temperature $T$ is much less than the exchange energy $J$, the decay constant is proportional to $T^n$, where $n$ is the order of the Jordan-Wigner fermion dispersion, namely $n=3$ for the XY chain proper and $n=2$ in the presence of a longitudinal magnetic field. When $T \gtrsim J$, the decay constant scales linearly with $T$. On the other hand, as this model possesses no inherent dissipation, the echo signal saturates to finite value instead of fading away as the waiting time $t_w\to \infty$. These results corroborate the physical picture presented in the previous work, and illustrate the 2DCS' capability in detecting subtle dynamical properties of optical excitations in the context of spin chains.

The rest of this work is organized as follows. In Sec.~\ref{sec:setup}, we define the problem. After describing the numerical method in Sec.~\ref{sec:method}, we present our results in Sec.~\ref{sec:results}. Finally, we discuss a few open problems in Sec.~\ref{sec:discussion}.

%%%%%%%%%%%%%%%%%%%%
% Problem set up
%%%%%%%%%%%%%%%%%%%%
\section{Problem set up \label{sec:setup}}

The Hamiltonian for the $S=1/2$ XY spin chain is given by~\cite{Lieb1961}:
\begin{align}
\label{eq:h_xy}
H = -\frac{J}{2}\sum^{N-1}_{j=1} (S^+_j S^-_{j+1} + H.C.)-B\sum^N_{j=1}S^z_j.
\end{align}
Here, $S^{\pm,z}_j$ are $S=1/2$ spin operators on site $j$. We employ the open boundary condition. $J$ is the exchange constant, which is chosen to be positive (ferromagnetic) without loss of generality. The antiferromagnetic ($J<0$) case is related to the present case by a staggered gauge transformation, namely rotating the spins on even sites by $\pi$ with respect to the $z$ axis. $B$ is a longitudinal field. 

The system hosts a TLL when $B<J$. In this work, we consider the regime $B\ll J$. The highly magnetized TLL ($B\lesssim J$) and the polarized phase ($B>J$) are beyond the scope of this work as these regimes have very different physics.

The quantities of central interest are the third order nonlinear magnetic susceptibilities $\chi^{(3)}_{+--+}$ and $\chi^{(3)}_{-++-}$. Both exhibit the photon echo described in Sec.~\ref{sec:intro} as per the analysis in ~\cite{li2021photon}. As these two susceptibilities show very similar behaviors, we only present results concerning $\chi^{(3)}_{+--+}$ for brevity. To set the stage, we define the following response function through the Kubo formula:
\begin{align}
\label{eq:tilde_chi3_def}
\tilde{\chi}_{+--+}^{(3)}(3,2,1,0) = i^3 \Theta(t_1)\Theta(t_2-t_1)\Theta(t_3-t_2) \nonumber\\ 
\times \langle [S_{j_3}^+(t_3), [S_{j_2}^-(t_2), [S_{j_1}^-(t_1), S_{N/2+1}^+(0)]]] \rangle. 
\end{align}
Here, $0\sim 3$ are shorthand notations for spacetime coordinates $(t_0, j_0)\sim (t_3,j_3)$. We have utilized the space and time translation symmetry to shift $t_0$ to $0$ and $x_0$ to the center of the chain $N/2+1$. Since the wavelength of the THz electromagnetic wave is much larger than the relevant microscopic length scales, we assume the spins couple to the THz field homogeneously. Therefore, the optical response of the system is obtained by summing over the lattice sites. It is convenient to define:
\begin{align}
\label{eq:chi3_def}
\chi^{(3)}_{+--+}(q|t, t_w, \tau) &=  \sum_{j_1}\sum_{j_2}\sum_{j_3} \text{e}^{-iq(j_1 + j_2 - j_3)} \nonumber \\ 
&\times \tilde{\chi}_{+--+}^{(3)}(3,2,1,0).
\end{align}
Here, $t = t_3-t_2$, $t_w = t_2-t_1$, and $\tau = t_1$ are respectively the gating time, the waiting time, and the delay time. $q=0$ for the ferromagnetic chain, and $q=\pi$ for the antiferromagnetic chain owing to the aforementioned staggered gauge transformation. Note the cases with other values of $q$ are potentially relevant to XY chain with Dzyaloshinskii-Moriya interactions~\cite{Gangadharaiah2008,Povarov2022}. In this work, we solely consider $q=0$ and $\pi$. 

$\chi^{(3)}_{+--+}$ and $\chi^{(3)}_{-++-}$ can be directly probed by 2DCS using THz pulses with appropriate circular polarizations. Their contributions are also visible with linearly polarized THz pulses since other third order susceptibilities do not exhibit echo signal. This feature eases up on the experimental conditions.

Finally, we obtain the two-dimensional spectrum by performing a Fourier transform of $\chi^{(3)}_{+--+}(q|t, t_w, \tau)$ with respect to $t$ and $\tau$:
\begin{align}
\label{eq:chi3_fft}
\chi^{(3)}_{+--+}(q|\omega_t, t_w, \omega_{\tau}) &= \int^\infty_0 dt \int^\infty_0 d\tau \,e^{i(\omega_t t+\omega_\tau \tau)}
\nonumber\\ 
& \times \chi^{(3)}_{+--+}(q|t, t_w, \tau). 
\end{align}
$\omega_t$ and $\omega_\tau$ are respectively the frequency variables conjugate to the gating time $t$ and the pulse delay time $\tau$.

\section{Method \label{sec:method}}

In this section, we sketch our numerical method for calculating Eq.~\eqref{eq:tilde_chi3_def} and Eq.~\eqref{eq:chi3_def}. Our procedure is similar in spirit to earlier works~\cite{Derzhko1998,Maeda2003}, namely expressing the spin correlation functions in terms of the non-local correlation functions of Jordan-Wigner fermions. The main improvement lies in the manner in which the non-local fermion correlation functions are treated. In the preceding works, they are computed in terms of Pfaffians of fermion Green's functions. Here, we recast them as \emph{determinants}~\cite{McCoy1971,Colomo1993,Zvonarev2009}. The resulted expressions are more compact in form and faster to evaluate, which speeds up the numerical calculation of the four-point spin correlation functions considerably.

In the first step, we diagonalize the $S=1/2$ XY spin chain by way of the Jordan-Wigner transformation,
\begin{align}
S_j^+ = \prod_{n<j} (1-2c_n^\dagger c^{\phantom\dagger}_n) c_j^\dagger, \quad
S_j^z = c_j^\dagger c^{\phantom\dagger}_j - 1/2,
\end{align}
where $c^\dagger_j$ and $c_j$ are fermion creation and annihilation operators. After the transformation, Eq.~\eqref{eq:h_xy} assumes the form of free fermion chain:
\begin{align}
\label{eq:h_f}
H &= -\frac{J}{2}\sum^{N-1}_{j=1} (c_j^\dagger c^{\phantom\dagger}_{j+1} +H.C.) - B\sum^N_{n=1} c_j^\dagger c_j
\nonumber \\
& = \sum_{k} \epsilon(k) c^\dagger_k c^{\phantom\dagger}_k.\quad (N\to\infty)
\end{align}
In the second line, we have taken the thermodynamic limit and switched to the momentum space. $\epsilon(k)$ is the dispersion relation of the Jordan-Wigner fermions:
\begin{subequations}
\label{eq:dispersion}
\begin{align}
\epsilon(k) = -J\cos k - B. 
\end{align}
The ground state is thus given by the Fermi sea. The dispersion near the Fermi point $k_F$ can be approximated as:
\begin{align}
\epsilon(k) \approx \left\{\begin{array}{cc}
v_F(k-k_F)-\frac{J}{6}(k-k_F)^3 & (B=0) \\
v_F(k-k_F)- \frac{B}{2}(k-k_F)^2 & (B\neq 0) 
\end{array}\right. ,
\end{align}
\end{subequations}
with the Fermi velocity $v_F = \sqrt{J^2-B^2}$. We see that leading correction to the linear dispersion relation is cubic (quadratic) in the absence (presence) of the longitudinal field. We shall see their different impacts on the photon echo in Sec.~\ref{sec:results}.

It is convenient for later purposes to define two sets of fermion Green's functions:
\begin{subequations}
\label{eq:Green}
\begin{align}
G^<_{j_1j_2}(t_1,t_2) &= \langle c^\dagger_{j_1}(t_1) c_{j_2}(t_2)\rangle.
\\
G^>_{j_1j_2}(t_1,t_2) &= -\langle c_{j_2}(t_2) c^\dagger_{j_1}(t_1)\rangle.
\end{align}
\end{subequations}
Note their definitions differ from the usual ``$G$-lesser" and ``$G$-greater" functions by a factor of $i$. They can be easily calculated from Eq.~\eqref{eq:h_f}.

In the next step, we compute the four-point response function Eq.~\eqref{eq:tilde_chi3_def} by expanding the nested commutators. This process yields a few four-point spin correlation functions. These, in turn, are expressed as non-local correlation functions of the Jordan-Wigner fermions. We recast these non-local correlation functions as determinants of fermion Green's functions $G^{>}$ and $G^{<}$. 

The starting point is the following formula for a general multi-point fermion correlation function:
\begin{subequations}
\label{eq:Pf_block}
\begin{align}
\left\langle \prod_{n=1}^{M} c^\dagger_{\alpha_n} c^{\phantom\dagger}_{\beta_n} \right\rangle = \det Y,
\end{align}
where $\alpha_n$ and $\beta_n$ are arbitrary labels of fermion modes. Note the product is understood as an ordered product with $n=1$ on the leftmost position and $n=M$ on the rightmost position. $Y$ is an $M \times M$ matrix whose entries are fermion Green's functions: 
\begin{align}
\label{eq:Y}
Y_{ij} =\left\{ \begin{array}{cc}
\langle c^\dagger_{\alpha_i} c^{\phantom\dagger}_{\beta_j} \rangle & i \le j, \\ 
-\langle c^{\phantom\dagger}_{\beta_j} c^\dagger_{\alpha_i} \rangle & i > j.
\end{array}\right. . 
\end{align}
\end{subequations}
Eq.~\eqref{eq:Pf_block} can be derived straightforwardly by using the Wick theorem.

As a preparation for later calculations, we consider the expectation value of a non-local operator that resembles the Jordan-Wigner string:
\begin{align}
\label{eq:str_exp} 
\left\langle \prod_{n=1}^M (1-2c^\dagger_{\alpha_n} c^{\phantom\dagger}_{\beta_n}) \right\rangle
&= \sum_{E} (-2)^{|E|} \left \langle \prod_{i\in E} c^\dagger_{\alpha_i}c^{\phantom\dagger}_{\beta_i} \right\rangle \nonumber \\ 
&=\sum_{E}\det(-2Y_{E, E}) \nonumber \\ 
&= \det(\text{I} - 2Y).
\end{align}
Here, $E$ is an subset of the $M$-element index set $\{1,2,\cdots M\}$. The summation is over all possible subsets including the empty set. $|E|$ denotes the number of elements of $E$. In the first line, we have expanded the product. In the second line, we have used Eq.~\eqref{eq:Pf_block}. Here, $Y_{E,E}$ is a submatrix of $Y$ (Eq.~\eqref{eq:Y}), whose rows and columns are chosen from the set $E$. Its determinant $\det Y_{EE}$ is known as the \emph{principal minor} of the matrix $Y$. In the last line, we employ the summation formula for the principal minors~\cite{Meyer2000}.

We proceed to calculate the spin correlation functions. It is sufficient for our purpose to illustrate the procedure for the two-point spin correlation functions. The four-point spin correlation functions is obtained in the same vein. Consider:
\begin{align}
\label{eq:exp_+-}
\langle S^+_{j_1}(t_1)S^-_{j_2}(t_2) \rangle &= \langle \prod_{n=1}^{j_1-1}(1-2c_n^\dagger(t_1)c_n(t_1))c_{j_1}^\dagger(t_1) \nonumber\\
& \times \prod_{m=1}^{j_2-1}(1-2c_m^\dagger(t_2)c_m(t_2))c_{j_2}(t_2) \rangle. 
\end{align}
We observe that the above expectation value, although resembles Eq.~\eqref{eq:str_exp}, is not exactly identical to it. Specifically, the operator $c_{j_1}$ misses a creation operator ($c^\dagger$) partner, and, likewise, $c_{j_2}$ misses an annihilation operator ($c$) partner. We remedy this situation by introducing auxiliary fermion operators $\psi$, $\psi^\dagger$, which we take to be algebraically independent from $c$ fermions, i.e. $\psi$ and $c$ modes mutually anti-commute. We further assume the $\psi$ mode is unoccupied, $\langle \psi^\dagger \psi\rangle = 0$. Using these, we rewrite Eq.~\eqref{eq:exp_+-} as:
\begin{widetext}
\begin{subequations}
\begin{align}
\label{eq:exp_+-_str_form}
\langle S^+_{j_1}(t_1)S^-_{j_2}(t_2) \rangle & = \frac{1}{4} \left\langle \prod_{n=1}^{j_1-1}(1-2c_n^\dagger(t_1)c_n(t_1)) (1-2c_{j_1}^\dagger(t_1)\psi) \prod_{m=1}^{j_2-1}(1-2c_m^\dagger(t_2)c_m(t_2)) (1-2\psi^\dagger c_{j_2}(t_2)) \right\rangle
\nonumber\\
& -\frac{1}{4} \left\langle \prod_{n=1}^{j_1-1}(1-2c_n^\dagger(t_1)c_n(t_1)) \prod_{m=1}^{j_2-1} (1-2c_m^\dagger(t_2)c_m(t_2)) \right\rangle
\nonumber \\ 
& = \frac{1}{4}(\det (I-2A) -\det (I-2B)). 
\end{align}
The first equality follows from the definition of the $\psi$ fermions. The second equality follows from Eq.~\eqref{eq:str_exp}. The matrices $A$ and $B$ are given by:
\begin{align}
A= 
\begin{pmatrix}
G^<_{1:j_1-1, 1:j_1-1}(t_1,t_1) & 0_{(j_1-1)\times 1} & G^<_{1:j_1-1, 1:j_2-1}(t_1, t_2) & G^<_{1:j_1-1,j_2}(t_1,t_2) \\
G^<_{j_1, 1:j_1-1}(t_1,t_1) & 0 & G^<_{j_1,1:j_2-1}(t_1,t_2) & G^<_{j_1,j_2}(t_1,t_2) \\
G^>_{1:j_2-1,1:j_1-1}(t_2,t_1) & 0_{(j_2-1)\times1} & G^<_{1:j_2-1,1:j_2-1}(t_2,t_2) & G^<_{1:j_2-1:j_2}(t_2,t_2) \\
0_{1\times (j_1-1)} & -1 & 0_{1\times (j_2-1)} & 0
\end{pmatrix};
\end{align}
and
\begin{align}
B=
\begin{pmatrix}
G^<_{1:j_1-1, 1:j_1-1}(t_1,t_1) & G^<_{1:j_1-1, 1:j_2-1}(t_1, t_2) \\  
G^>_{1:j_2-1, 1:j_1-1}(t_1, t_2) & G^<_{1:j_2-1, 1:j_2-1}(t_2, t_2)
\end{pmatrix}.
\end{align}
\end{subequations}
Here, the lesser and greater Green's functions are defined in Eq.~\eqref{eq:Green}. $G^{<}_{i:j, k:l}$ denotes the submatrix of $G^<$ whose rows and columns range from $i$ to $j$ and $k$ to $l$, respectively. $G^{>}_{i:j, k:l}$ is defined in the same vein. Substituting the above into Eq.~\eqref{eq:exp_+-_str_form}, and performing the Laplace expansion of $\det(I-2A)$ along the $j_1$-th column and the last row yields:
\begin{align}
\label{eq:exp_+-_det}
\langle S^+_{j_1}(t_1)S^-_{j_2}(t_2) \rangle &=\frac{(-1)^{j_2}}{2}
\det
\begin{pmatrix}
[I-2G^<(t_1,t_1)]_{1:j_1, 1:j_1-1} & -2G^<_{1:j_1, 1:j_2}(t_1,t_2) \\ 
-2G^>_{1:j_2-1, 1:j_1-1}(t_1,t_2) & [I-2G^<(t_2,t_2)]_{1:j_2-1,1:j_2}
\end{pmatrix}.
\end{align}

The calculation of the four-point spin correlation function parallels that of the two-point spin correlation function but is more involved. We only present the result for brevity. All four point correlation function can be cast in a fairly regular form:
\begin{align}
\label{eq:4point_det}
& \langle S^{\mu_1}_{j_1}(t_1)S^{\mu_1}_{j_2}(t_2)S^{\mu_3}_{j_3}(t_3)S^{\mu_4}_{j_4}(t_4) \rangle = \frac{(-1)^{j_2+j_4}}{4} 
\nonumber \\ 
&\times \det
\begin{pmatrix}
[I-2G^<(t_1,t_1)]_{J_1,J'_1} 
& -2G^<(t_1, t_2)_{J_1, J'_2} 
& -2G^<(t_1, t_3)_{J_1, J'_3}
& -2G^<(t_1, t_4)_{J_1, J'_4} \\ 
-2G^>(t_1, t_2)_{J_2,J'_1}
& [I-2G^<(t_2,t_2)]_{J_2,J'_2} 
& -2G^<(t_2, t_3)_{J_2,J'_3}
& -2G^<(t_2, t_4)_{J_2,J'_4}\\ 
-2G^>(t_1, t_3)_{J_3,J'_1}
& -2G^>(t_2, t_3)_{J_3,J'_2}
& [I-2G^<(t_3,t_3)]_{J_3, J'_3}
& -2G^<(t_3, t_4)_{J_3 J'_4}\\ 
-2G^>(t_1, t_4)_{J_4.J'_1}
& -2G^>(t_2, t_4)_{J_4,J'_2}
& -2G^>(t_3, t_4)_{J_4,J'_3}
& [I-2G^<(t_4,t_4)]_{J_4,J'_4}
\end{pmatrix}. 
\end{align}
\end{widetext}
Here, the greek indices $\mu_{1,2,3,4} = \pm$ refer to the raising/lowering type of the spin operator. $J_{1,2,3,4}$ and $J'_{1,2,3,4}$ are index sets. $J_{i} = \{1,2,\cdots j_i\}$ ($\{1,2,\cdots j_i-1\}$) when $\mu_i = +$ ($\mu_i = -$). Likewise, $J'_{i} = \{1,2,\cdots j_i-1\}$ ($\{1,2,\cdots j_i\}$) when $\mu_i = +$ ($\mu_i = -$). 

The final step is carrying out the summation over the lattice sites $j_{1,2,3}$ in Eq.~\eqref{eq:chi3_def}. Instead of summing over all sites, we reduce the computational work load by utilizing causality, namely $\tilde{\chi}_{+--+}^{(3)}(0, 1, 2, 3)$ falls off exponentially outside the light cone about the center of the chain. Therefore, we may restrict the summation over $j_{1,2,3}$ to $\pm (Jt_{1,2,3}+R)$, where $Jt$ is the radius of the light cone at time $t$, and $R$ represents the size of a small interval outside the cone. We choose $R=5$, which yields a relative error on the order of $10^{-3}$. Furthermore, we reduce the work load by half by using the inversion symmetry with respect to the center of the chain. 

We close this section by commenting on the computational complexity. For a chain with $N$ sites, computing a four-point spin correlation function requires evaluating a single determinant of typical size $O(N)$, whose complexity is $O(N^3)$. Taking the triple lattice summation into account, the complexity for computing the third-order susceptibility is thus $O(N^6)$. This scaling limits the numerically accessible system size, and, as a result, the simulation time and temperature. With the various improvements described in this section, we are able to compute the full two-dimensional spectrum of systems up to $N=157$ with $O(10^6)$ CPU hours. For calculating the magnitude of photon echo signal, we can reach system size up to $N = 357$ by further restricting the lattice summation to sites close to the lensing configurations (Fig.~\ref{fig:sche}(b)).

\begin{figure*}
\includegraphics[width = 2\columnwidth]{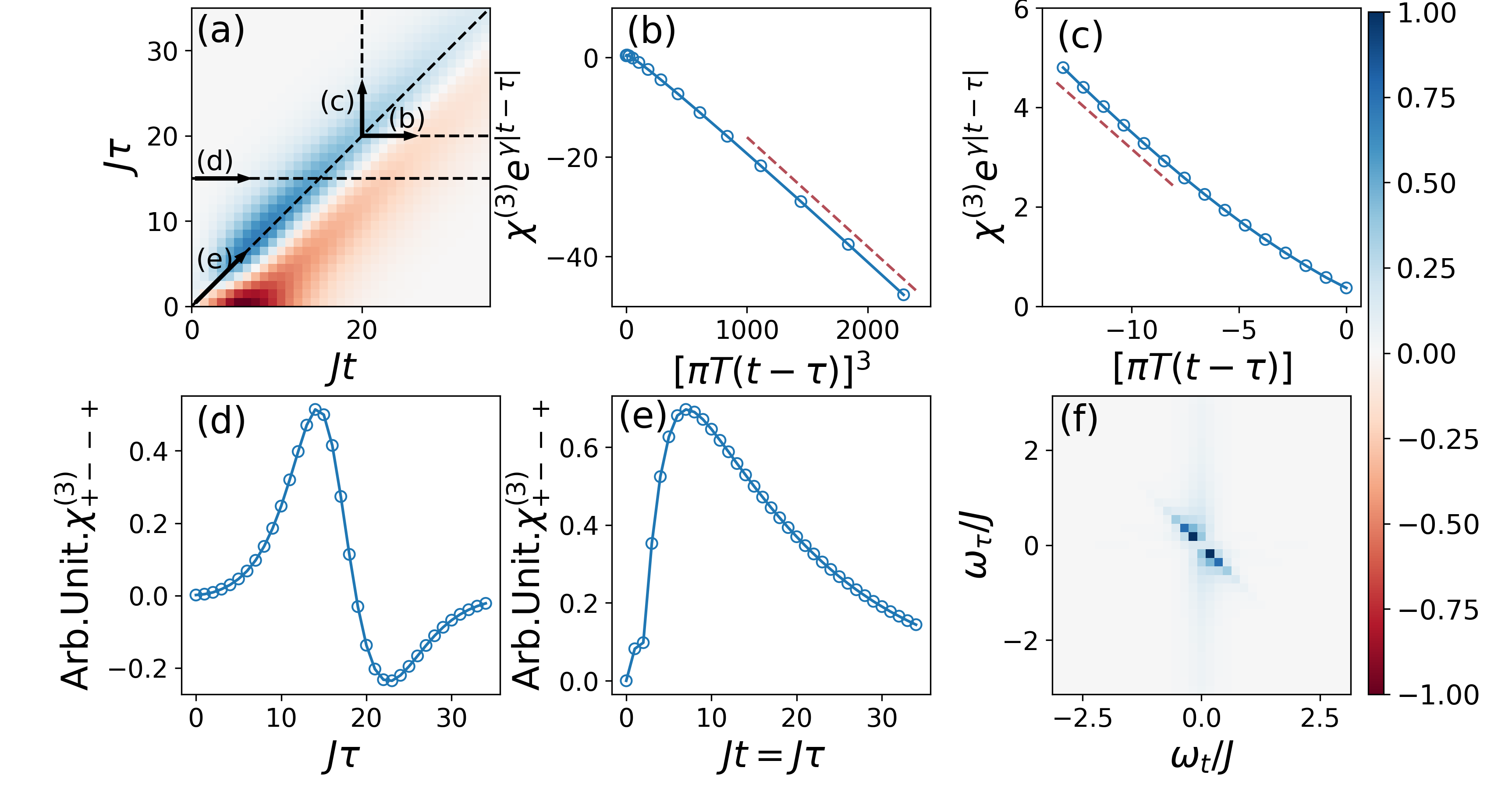}
\caption{(a) Nonlinear magnetic susceptibility $\chi_{+--+}^{(3)}$ as a function of $t$ and $\tau$ for the ferromagnetic ($q=0$ in Eq.~\ref{eq:chi3_def}) chain at $T/J=0.3$ and in zero field. The waiting time $t_w = 0$. The data are rescaled such that its maximum magnitude equals to $1$. (b) A scan of the data with constant $\tau$ as indicated by the arrow ``(b)" in panel (a). The red dashed line delineates the expected asymptotic behavior. (c) A constant $t$ scan of the data as indicated by the arrow ``(c)" in panel (a). The red dashed line delineates the expected asymptotic behavior. (d) A constant $\tau$ scan of the data as indicated by the arrow ``(d)" in panel (a), which shows the full profile of the photon echo.  (e) The data along the diagonal direction of the $t$-$\tau$ plane as indicated by the arrow ``(e)" in panel (a). (f) The complex modulus of the two dimensional spectrum. The data are rescaled similarly to (a).}
\label{fig:FM_2DCS}
\end{figure*}

\begin{figure*}
\includegraphics[width = 2\columnwidth]{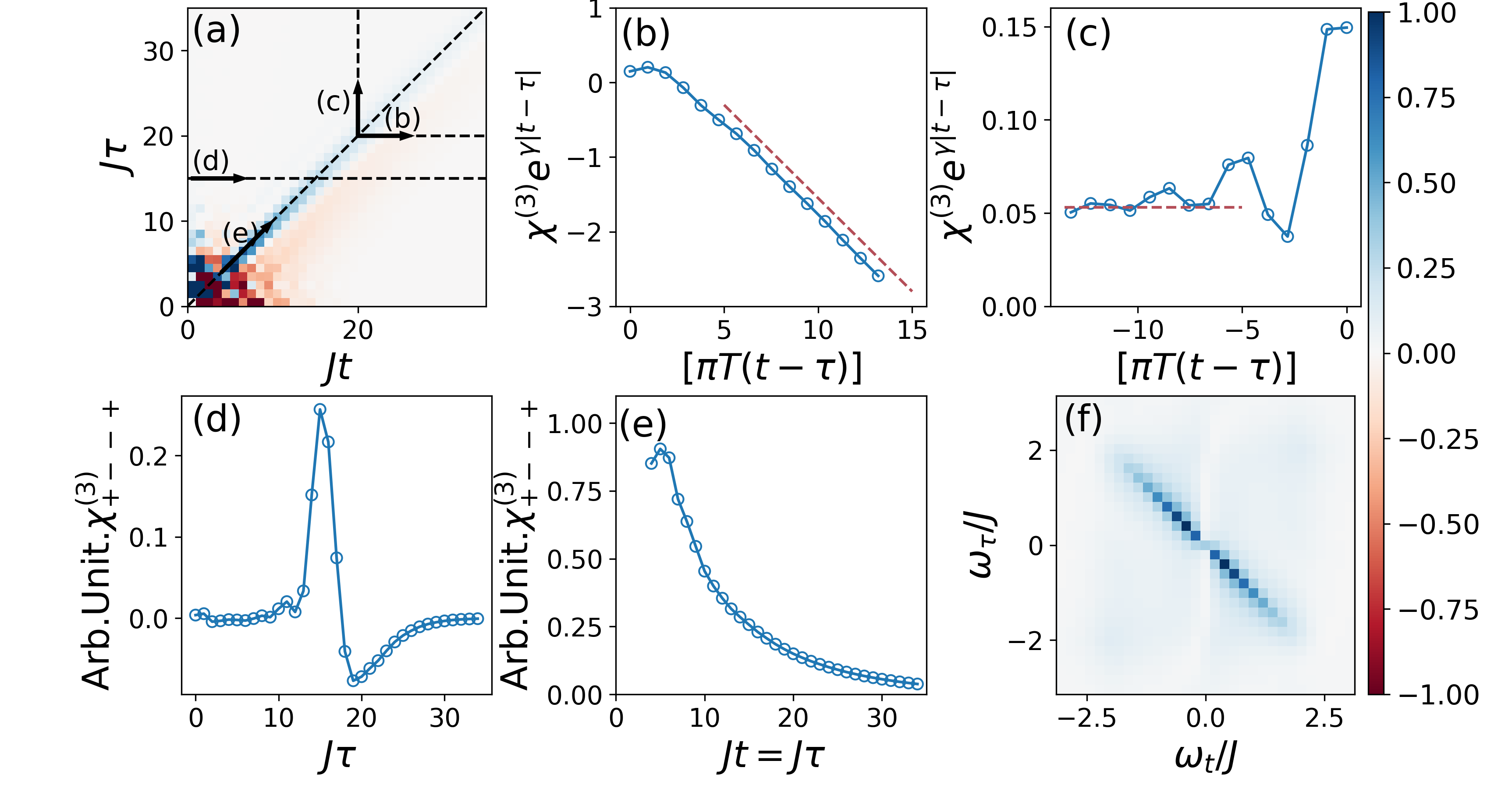}
\caption{The same as Fig~\ref{fig:FM_2DCS} but for antiferromagnetic ($q=\pi $ in Eq.~\eqref{eq:chi3_def}) chain.}
\label{fig:AFM_2DCS}
\end{figure*}

\section{Results \label{sec:results}}

Fig.~\ref{fig:FM_2DCS}(a) shows the numerically calculated nonlinear magnetic susceptibility $\chi_{+--+}^{(3)}$ as a function of $t$ and $\tau$ for the ferromagnetic chain ($q=0$ in Eq.~\ref{eq:chi3_def}) at $T/J=0.3$ and $B=0$. We set waiting time $t_w =0$. Note the susceptibility is a real number in this case. The photon echo signal appears as the feature running along the diagonal of the $t$-$\tau$ plane. Taking a cut with constant  $\tau$ reveals the structure of the echo signal (Fig.~\ref{fig:FM_2DCS}(d), with $J \tau = 15$). We see the nonlinear response reaches the maximum at $t\approx \tau$, which is the characteristic of the photon echo. Our previous analysis predicts the following asymptotic behavior for the photon echo signal~\cite{li2021photon}:
\begin{align}
\chi_{+--+,q=0}^{(3)} \sim  \left\{ \begin{array}{cc}
(t-\tau)^3 e^{-\gamma |t-\tau|} & (t \gg \tau \gg 0) \\
(t-\tau) e^{-\gamma |t-\tau|} & (\tau \gg t \gg 0)
\end{array}\right. ,
\end{align}
where $\gamma = \pi T/2$. The numerical results confirm these predictions (Fig.~\ref{fig:FM_2DCS}(b)(c)). 

For the ideal TLL where the fractional excitations neither disperse nor dissipate, our previous analysis shows that the echo persists along the diagonal direction of the $t$-$\tau$ plane, i.e. the echo does not depend on $\tau$. Here, the diagonal feature in Fig.~\ref{fig:FM_2DCS}(a) associated with the photon echo gradually fades away at large $\tau,t$. The decay of the photon echo signal is best illustrated by taking a cut of Fig.~\ref{fig:FM_2DCS}(a) along the diagonal direction of the $t$-$\tau$ plane (Fig~\ref{fig:FM_2DCS}(e)). After the initial rise, the echo signal decreases exponentially as $\tau = t$ increases. This information is also encoded in the two-dimensional spectrum (Fig~\ref{fig:FM_2DCS}(f)), where the photon echo is manifest as a pair of peaks on the $\omega_t$-$\omega_\tau$ plane. The anti-diagonal width of photon echo peaks is inversely proportional to the decay time of the echo signal, whereas the diagonal width of the peaks scales with temperature linearly.

We attribute the decay of the photon echo signal to the nonlinearity of the dispersion relation of the Jordan-Wigner fermions (Eq.~\eqref{eq:dispersion}). A key step in lensing is the refocusing of the excitation world lines by the second and third pulses (Fig.~\ref{fig:sche}(b)). This process occurs because acting the spin raising/lowering operator on the fractional excitations reverses their direction of motion and changes the topological charge they carry~\cite{li2021photon}. However, due to the aforementioned nonlinearity, a fractional excitation wave packet disperses as it travels through the system. As a result, the refocusing is no longer prefect, which suppresses the lensing process. We shall discuss further on the decay of the photon echo momentarily.

Having discussed the ferromagnetic chain, we turn to the antiferromagnetic case ($q=\pi $ in Eq.~\eqref{eq:chi3_def}). Fig~\ref{fig:AFM_2DCS}(a) shows the nonlinear magnetic susceptibility $\chi_{+--+}^{(3)}$ as a function of $t$ and $\tau$ with the same set of model parameters. Compared to the ferromagnetic case, the photon echo signal is disguised at early time by other nonlinear responses; nonetheless, it is clearly visible at late times. Fig.~\ref{fig:AFM_2DCS}(d) shows the profile of the photon echo along a cut with constant $J\tau = 15$. We note its overall span on the time axis is shorter than that of the ferromagnetic case (Fig.~\ref{fig:FM_2DCS}(d)). 

Our previous work predicts the following asymptotic behavior for the photon echo in the antiferromagnetic case~\cite{li2021photon}:
\begin{align}
\chi_{+--+,q=\pi}^{(3)} \sim  \left\{ \begin{array}{cc}
(t-\tau) e^{-\gamma |t-\tau|} & (t \gg \tau \gg 0) \\
e^{-\gamma |t-\tau|} & (\tau \gg t \gg 0)
\end{array}\right. ,
\end{align}
where $\gamma$ is the same as before~\footnote{Note, however, the constant $\gamma$ taking the same value for the ferro- and antiferromagnetic chain is a special feature of the XY chain, or TLL with $K=1$. See \onlinecite{li2021photon} for the expression of $\gamma$ for general $K$.}. We find good agreement between the numerical results and the prediction for $t>\tau$ (Fig.~\ref{fig:AFM_2DCS}(b)) but much less so for $t<\tau$ (Fig.~\ref{fig:AFM_2DCS}(c)). The small magnitude of the signal for the latter case makes it challenging to numerically ascertain the asymptotic behavior. Similar to the ferromagnetic case, the photon echo decreases exponentially with $\tau$ (Fig.~\ref{fig:AFM_2DCS}(e)).

Fig~\ref{fig:AFM_2DCS}(f) shows the two dimensional spectrum. Note we have masked out the data with $Jt<3$ and $J\tau<3$ in the Fourier transform to enhance the photon echo peaks. We see that the photon echo peaks are more extended in the frequency plane compared with the ferromagnetic chain, which is directly related to its shorter span on the time axis (Fig.~\ref{fig:FM_2DCS}(d)) relative to the ferromagnetic case.

\begin{figure}
\includegraphics[width = \columnwidth]{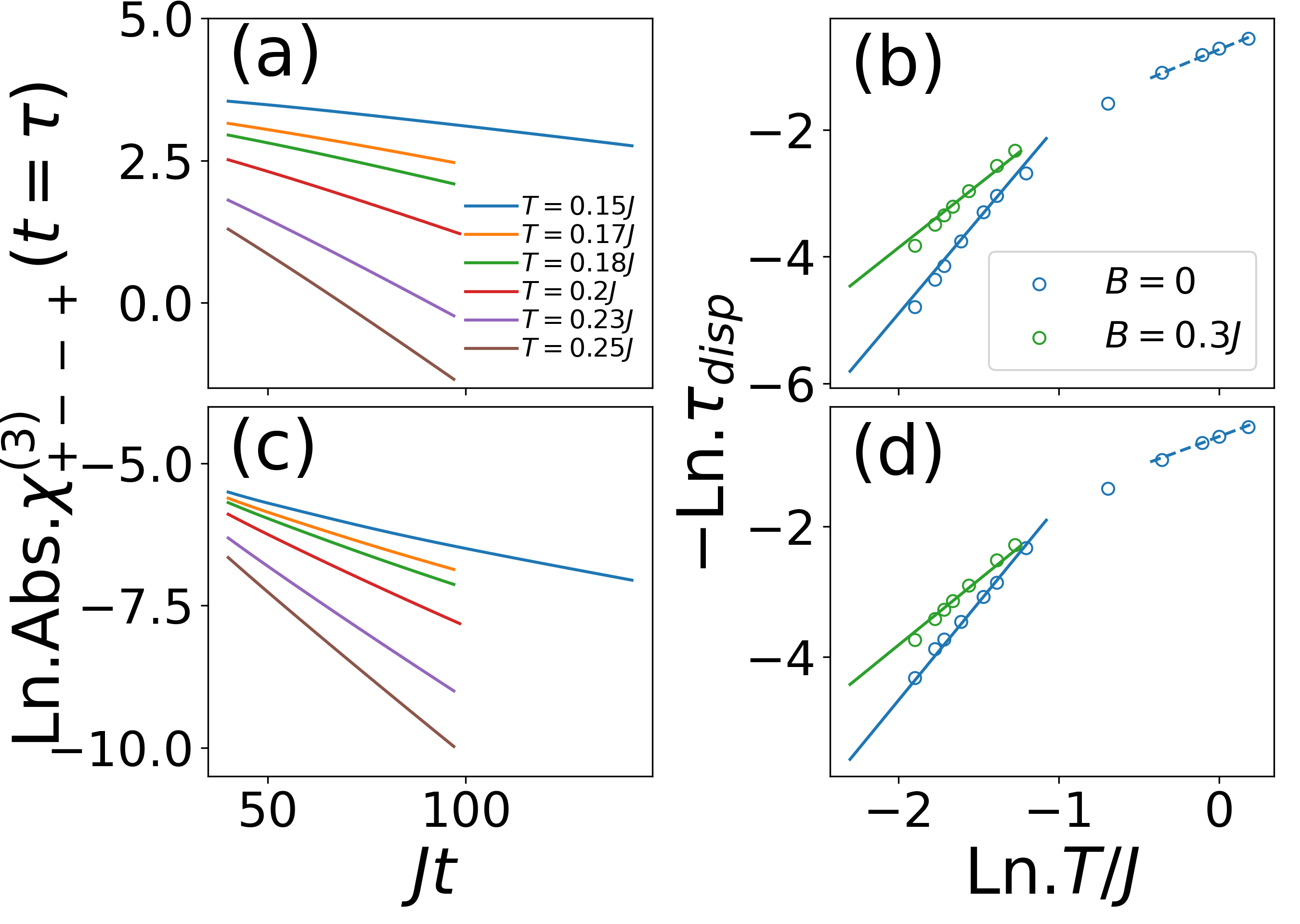}
\caption{(a) The behavior of $\chi_{+--+}^{(3)}$ at $t=\tau$ for the ferromagnetic chain at various temperatures and in zero field. The waiting time $t_w = 0$. (b) The decay time $\tau_\mathrm{disp}$ as a function of temperature. The blue and green open circles represent the case with $B=0$ and $B=0.3J$, respectively. The blue solid line, the blue broken line, and the green solid line delineate respectively the power law $T^\alpha$ with exponents $\alpha=3$, $1$, and $2$, respectively. (c)(d) Same as (a)(b) but for antiferromagnetic chain.}
\label{fig:tau_disp}
\end{figure}

\begin{figure}
\includegraphics[width = \columnwidth]{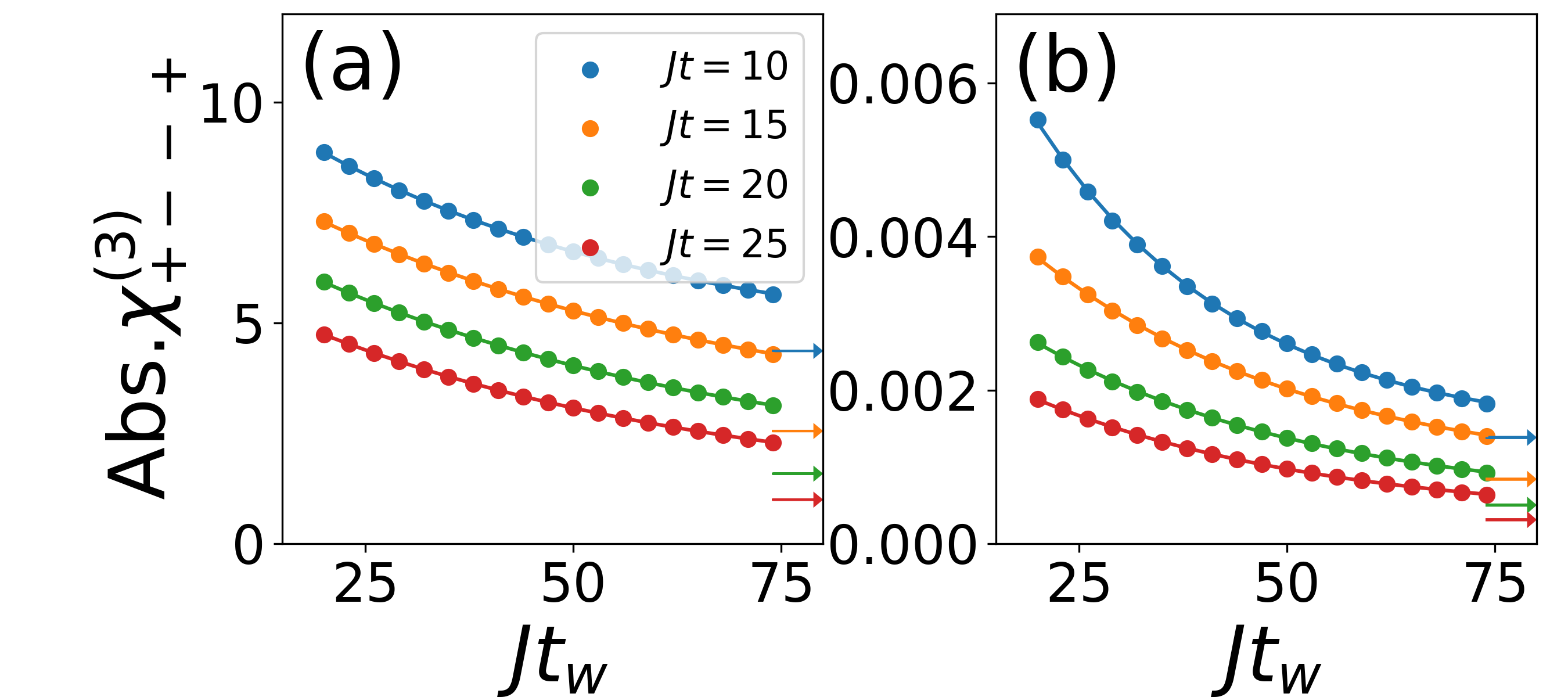}
\caption{The nonlinear magnetic susceptibility $\chi^{(3)}_{+--+}$ as a function of $t_w$ at various values of $t=\tau$ for the ferromagnetic (a) and antiferromagnetic (b) chain. The temperature $T/J = 0.25$, and the magnetic field $B = 0$. Numerical data are denoted as filled circles. Solid lines are fitting of the data to exponential functions. Arrows mark the extrapolated value of $\chi^{(3)}_{+--+}$ at $t_w\to \infty$. }
\label{fig:tw_dependence}
\end{figure}

We now investigate the decay of the photon echo in more detail. Fig.~\ref{fig:tau_disp}(a)(c) (solid lines) show the value of $\chi_{+--+}^{(3)}$ along the diagonal direction ($\tau = t$) in the $t$-$\tau$ plane at various temperatures for both the ferromagnetic and the antiferromagnetic chains. For both cases, and at all numerically accessible temperatures, we find the echo signal decreases exponentially at late time, i.e. 
\begin{align}
\chi^{(3)}_{+--+}(t=\tau) \sim e^{-\tau/\tau_\mathrm{disp}},
\end{align}
where $\tau_\mathrm{disp}$ is the decay time of the echo signal. Physically, we interpret $\tau_\mathrm{disp}$ as an effective life time of the fractional excitations created by the first pulse --- in the case of the ferromagnetic chain, they are a pair of spinons; in the case of the antiferromagnetic chain, they are a spinon and a Laughlin quasiparticle~\cite{Pham2000}. Their wave packets broaden as they propagate. Beyond the time $\tau_\mathrm{disp}$, they are virtually indistinguishable from the thermal fluctuations. Fig.~\ref{fig:tau_disp}(b)(d) (open circles) present the dependence of $\tau_\mathrm{disp}$ on temperature, extracted by fitting the $\chi^{(3)}_{+--+}(t=\tau)$ data to an exponential function. We identify two regimes: At low temperature $T \ll J$, $1/\tau_\mathrm{disp} \propto T^3$; at higher temperature $T\sim J$ or higher, we find $1/\tau_\mathrm{disp} \propto T$.

We may understand the low temperature $T^3$ scaling by a simple dimension counting argument. At the renormalization group (RG) fixed point, the Hamiltonian of TLL is that of free, relativistic bosons. At the fixed point, the lensing is perfect, and the photon echo does not decay. Now, the nonlinearity in the dispersion relation of the Jordan-Wigner fermions at $B=0$ gives rise to the following RG irrelevant perturbation to the fixed point Hamiltonian:
\begin{align}
H' =  \int dx\, & \lambda_+(\nabla\phi_R)^2(\nabla\phi_L)^2 
\nonumber\\
 + & \lambda_-[(\nabla\phi_R)^4+(\nabla\phi_L)^4)]. 
\end{align}
The explicit value of $\lambda_\pm$ are determined by the microscopic model~\cite{LUKYANOV1998533,Igor,Samokhin_1998}. Both $\lambda_{\pm}$ have dimension $2$ in unit of length. Given that they are RG irrelevant, we expect $1/\tau_\mathrm{disp}$ admits a perturbative expansion in $\lambda_\pm$. The dimensional analysis then reveals $1/\tau_\mathrm{disp}\sim \lambda_\pm T^3$ to the leading order.

This analysis immediately implies that $1/\tau_\mathrm{disp} \sim T^n$, where $n$ is the leading order of the Jordan-Wigner fermion dispersion. In the presence of longitudinal magnetic field, the leading order changes from $n=3$ to $n=2$ (Eq.~\eqref{eq:dispersion}). Thus, we expect a $T^2$ scaling in this case. Our numerical results show that $1/\tau_\mathrm{disp} \propto T^2$ for both ferromagnetic and antiferromagnetic chains, in agreement with the analysis (Fig.~\ref{fig:tau_disp}(b),(d)).

So far, we have fixed the waiting time $t_w = 0$. The dependence of the photon echo on $t_w$ also contains important information about the dynamics of the system. Fig.~\ref{fig:tw_dependence}(a) and (b) present $\chi_{+--+}^{(3)}$ as a function of $t_w$ with fixed $t=\tau$ for ferromagnetic chain and antiferromagnetic chain, respectively. We set the temperature $T/J = 0.25$ and magnetic field $B=0$. For both cases, we find the data can be well fit to an exponential function, $ A\exp(-\alpha t_w) + A' $. Crucially, we find $A'\neq 0$; i.e. the magnitude of the echo signal saturates at $t_w\to\infty$ instead of decreasing to zero.

This behavior is associated to the integrability of the $S=1/2$ XY spin chain. In a thermalizing system, so long as the response is not tied to any conservation law or spontaneously broken symmetry, the nonlinear response function \emph{must} tend to zero as $t_w\to \infty$ because the memory about the first two pulses is lost. Therefore, the fact that $\chi^{(3)}_{+--+}$ does \emph{not} decrease to zero is necessarily a consequence of the integrability. In our previous work~\cite{li2021photon}, we hypothesized that the saturation of the photon echo signal with $t_w$ is a feature of integrable spin chains. Our results seem to support this idea.

\section{Discussion \label{sec:discussion}}

In this work, we have analyzed the photon echo of the $S=1/2$ XY spin chain. Through numerical calculations, we show that the photon echo signal decays with increasing pulse delay time $\tau$, whereas it saturates as the waiting time $t_w\to\infty$. The former reflects the suppression of lensing due to the dispersion of the fractional excitation wave packets. The latter, on the other hand, is a manifestation of the integrability of the model. These results are in broad agreement with the physical picture presented in the previous work. Furthermore, the numerically extracted asymptotic behavior of the echo signal is also quantitatively consistent with the prediction.

Our numerical calculation reveals that the decay rate of the photon echo (with the delay time $\tau$) scales with the temperature $T$ as $T^n$, where $n$ is the order of the leading nonlinearity in the dispersion relation of the Jordan-Wigner fermions. The nonlinearity in the fermion dispersion relation, and its physical consequences, are the central topics of the nonlinear Luttinger liquid theory~\cite{Imambekov2012,Imambekov2009}. Although a formidable task, it would be illuminating to treat the present problem analytically by using this approach. 

Adding \emph{generic} longitudinal exchange interactions to the $S=1/2$ XY chain breaks the integrability of the model. As a result, the fractional excitations now disperse as well as dissipate, which leads to additional decay of the echo signal with respect to the delay time $\tau$. In the bosonization language, this perturbation is amount to an umklapp term in the bosonized Hamiltonian. Comparing the RG eigenvalue of the umklapp term $2-4K$ to that of the dispersion term $-2$, we expect that the dispersion-induced decay remains the dominant mechanism when $K>1$~\cite{Igor}. Outside this regime, a systematic treatment of dissipation is necessary to clarify its impact.

Finally, we have seen that saturation of the echo signal with $t_w$ is associated with the integrability of the $S=1/2$ XY spin chain. We think that the integrability of a many-body system is manifest in a nonlinear response function is remarkable. In light of the studies on the out-of-time-order correlation functions~\cite{larkin1969quasiclassical,maldacena2016bound,maldacena2016remarks}, it is perhaps interesting to explore whether the photon echo could provide an in-depth characterization of quantum chaos, or the lack thereof, of many-body systems.

\begin{acknowledgments}
This work is supported by the National Natural Science Foundation of China (Grant No.~12250008, 11974396, and 12188101), by the National Key R\&D Program of China (Grant No.~2022YFA1403800), and by the Chinese Academy of Sciences through the Strategic Priority Research Program (Grant No.~XDB33020300), and the Project for Young Scientists in Basic Research (Grant No. YSBR-059).
\end{acknowledgments}

\bibliography{ver2.bib}

\end{document}